\documentclass{aa} %
\usepackage{graphicx}
\usepackage{txfonts}
\usepackage{color}
\usepackage{amssymb}
\usepackage{xspace}
\newcommand{\inte}{INTEGRAL\xspace}
\newcommand{\cyg}{Cyg~X-1\xspace}
\newcommand{\vfour}{V404~Cygni\xspace}
\newcommand{\maxi}{MAXI\xspace}
\newcommand{\bat}{Swift/BAT\xspace}
\newcommand{\ibis}{IBIS\xspace}

\newcommand{\rxte}{RXTE\xspace}
\newcommand{\jemx}{JEM-X\xspace}
\newcommand{\nicer}{NICER\xspace}

\newcommand\fn[1]{%
  \begingroup
  \renewcommand\thefootnote{}\footnote{#1}%
  \addtocounter{footnote}{-1}%
  \endgroup
}

\usepackage{subcaption}
\begin{document}

 \title{Unprecedentedly bright X-ray flaring in Cygnus~X-1 observed by \inte}

  \author{P. Thalhammer\inst{\ref{in:FAU}} \and 
  T. Bouchet\inst{\ref{in:UPU}}   \and 
  J. Rodriguez\inst{\ref{in:UPU}} \and 
  F. Cangemi\inst{\ref{in:APC}}   \and 
  K. Pottschmidt\inst{\ref{in:GSFC},\ref{in:CRESST}} $^{\dagger}$\and 
  D.A.\ Green\inst{\ref{in:AGCL}}   \and 
  L. Rhodes \inst{\ref{in:OX}}   \and
  C. Ferrigno\inst{\ref{in:GEN}}  \and   
  M.A.\ Nowak\inst{\ref{in:WUSL}}  \and
  V. Grinberg\inst{\ref{in:ESTEC}} \and
  T. Siegert\inst{\ref{in:WU}}  \and 
  P. Laurent\inst{\ref{in:UPU}}   \and
  I. Kreykenbohm\inst{\ref{in:FAU}} \and
  M. Perucho\inst{\ref{in:UV},\ref{in:OAUV}} \and
  J. Tomsick\inst{\ref{in:SSL}} \and
  C.~S\'anchez-Fern\'andez\inst{\ref{in:ESAC}} \and
  J. Wilms\inst{\ref{in:FAU}} 
  }
   \institute{%
     {Dr.\ Karl Remeis-Observatory, Friedrich-Alexander-Universit\"at
     Erlangen-N\"urnberg, Sternwartstr.~7, 96049 Bamberg, Germany \label{in:FAU}}
   \and
     {Universit\'e Paris Cit\'e, Universit\'e Paris-Saclay, CEA, CNRS,
     AIM, 91191 Gif-sur-Yvette, France  \label{in:UPU}}
   \and
     {APC, Universit\'e Paris Cit\'e, CNRS, CEA, Rue Alice Domont \&
     L\'eonie Duquet, 75013 Paris, France \label{in:APC}}
   \and
     {NASA Goddard Space Flight Center, Astrophysics Science Division,
     8800 Greenbelt Road, Greenbelt, MD 20771, USA \label{in:GSFC}}
   \and
     {CRESST and Center for Space Sciences and Technology, University
     of Maryland Baltimore County, 1000 Hilltop Circle, Baltimore, MD
     21250, USA \label{in:CRESST}}
   \and
     {Astrophysics Group, Cavendish Laboratory, J. J. Thomson Avenue,
      Cambridge CB3 0US, UK \label{in:AGCL}}
   \and
     {Astrophysics, Department of Physics, University of Oxford, Denys
     Wilkinson Building, Keble Road, Oxford OX1 3RH, UK \label{in:OX}}
   \and
     {University of Geneva, Department of Astronomy, Chemin d'Ecogia 16,
     1290 Versoix, Switzerland \label{in:GEN}}
   \and
     Washington University, MSC 1105-109-02, One Brookings Drive,  St.\,Louis, MO 63130-4899 \label{in:WUSL}
        \and
     {European Space Agency (ESA), European Space Research and
     Technology Centre (ESTEC), Keplerlaan 1, 2201 AZ Noordwijk, The Netherlands  \label{in:ESTEC}}
   \and
   {Julius-Maximilians-Universität Würzburg, Fakultät für Physik und Astronomie, Institut für Theoretische Physik und Astrophysik, Lehrstuhl für Astronomie, Emil-Fischer-Str 31, 97074 Würzburg, Germany  \label{in:WU}}
     \and
     {Departament d'Astronomia i Astrof\/isica, Universitat de Val\`encia, C/ Dr.\ Moliner, 50, 46100 Burjassot, València, Spain  \label{in:UV}}  
     \and 
     {Observatori Astron\`omic, Universitat de Val\`encia, C/ Catedr\`atic Jos\'e Beltr\'an, 46980 Paterna, Val\`encia, Spain  \label{in:OAUV}}
  \and
  {Space Sciences Laboratory, 7 Gauss Way, University of California, Berkeley, CA 94720-7450, USA \label{in:SSL}}
  \and
{European Space Agency (ESA), European Space Astronomy Centre (ESAC), Villafranca del Castillo, 28692 Madrid, Spain  \label{in:ESAC}}
   }
     
   \date{\today}

  \abstract{We study three
    extraordinarily bright X-ray flares originating from \cyg seen on
    2023~July~10 detected with \inte. The flares had a duration on the order of only ten
    minutes each, and within seconds reached a 1--100\,keV peak luminosity of
    $1.1\textrm{--} 2.6\times 10^{38}\,\mathrm{erg}\,\mathrm{s}^{-1}$. The
    associated \inte/IBIS count rate was about ${\sim}10\times$ higher
    than usual for the hard state. To our knowledge, this is the first
    time that such strong flaring has been seen in Cyg X-1, despite
    the more than 21\,years of \inte monitoring, with almost
    ${\sim}$20\,Ms of exposure, and the similarly deep monitoring with
    RXTE/PCA that lasted from 1997 to 2012. The flares were seen in
    all three X-ray and $\gamma$-ray instruments of \inte. Radio monitoring by the AMI Large Array with
    observations 6\,h before and 40\,h after the X-ray flares did
    not detect a corresponding increase in radio flux. The shape of
    the X-ray spectrum shows only marginal change during the flares,
    i.e., photon index and cut-off energy are largely preserved. The
    overall flaring behavior points toward a sudden and brief release
    of energy, either due to the ejection of material in an unstable jet or due to the interaction of the jet with the ambient clumpy stellar wind. }  \maketitle  
   
\section{Introduction} 
\cyg is one of the best studied stellar-mass black hole X-ray
binaries. Discovered in 1962 \citep{Bowyer65}, the black hole has a mass of $21.2\pm2.2\,M_\odot$\citep{MillerJones2021} and is in an almost circular 5.6\,d orbit with its donor, HDE 226868 \citep{Bolton72}, at a separation of 0.24\,AU and a distance of $2.22^{+0.18}_{-0.17}$\,kpc from us \citep{MillerJones2021}. Systematic long-term monitoring in the X-rays started around
1975 with Ariel V \citep{holt1979a}, was continued with Ginga
\citep{kitamoto2000a}, CGRO/BATSE \citep{ling1997}, and the
\rxte/ASM \citep{Grinberg2013} and is currently available from \maxi and
Swift/BAT. \cyg has also been the subject of detailed campaigns of pointed
observations, with missions such as \rxte
\citep{pottschmidt2003a,Wilms2006,Grinberg2014}, \inte
\citep[e.g.,][]{DelSanto2013,Cangemi2021_X1}, and \nicer \citep[e.g.,][]{konig2024}. In
total, almost half a century of X-ray monitoring has shed light on the
variability of \cyg on  timescales from milliseconds to months. 
\fn{$^\dagger$ deceased 17 June 2025}

\cyg is a persistent X-ray source that, similar to other BHBs, can be
found in two canonical states, which can be characterized through
their spectral and timing properties \citep[e.g.,][and references
  therein]{Grinberg2013,konig2024}. The ``hard state'' shows an X-ray
spectrum that can mainly be explained by Comptonization of soft seed photons by a hot electron plasma. In the ``soft
state'', thermal emission from the accretion disk dominates the
spectrum, although at least some Comptonized radiation is still
observed. Between the two main states \cyg transits through the so called ``intermediate state'', with changes between different states typically
happening on timescales of days to weeks. On short timescales of
milliseconds to minutes power spectra and other timing quantities
show characteristic state-dependent behavior that have been interpreted
as being due to a combination of variability in the accretion flow and in the
Comptonizing plasma \citep[][and references therein]{konig2024}.
Superimposed on and potentially distinct from this variability, flares
with a duration of seconds to minutes have been seen
\citep[e.g.,][]{wilms2007a}. Consistent with such flaring activity,
the power spectra sometimes display an additional noise component
below $\sim$0.01\,Hz \citep[e.g.,][]{vikhlinin1994}.

\cyg has also shown variable but persistent radio emission, which is
due to the presence of a radio jet \citep{Gallo2005,Rushton2012}. The
radio variability has been tracked during various years-long campaigns
with, e.g., the Ryle telescope, AMI, or MERLIN
\citep[e.g.,][]{fender2006,gleissner2004c,Rodriguez2015}. In
addition to a correlation between radio and hard X-ray flux in both
the hard and the soft state \citep{gleissner2004c,Zdziarski2020}, these
campaigns also showed the (rare) presence of radio flares which may be
similar to bubble ejection events in blazars
\citep[e.g.,][]{fender2006,pooley2017}. This includes a case where an
X-ray flare was followed by a radio flare $\sim$7\,min later
\citep{wilms2007a}.

X-ray flaring on intermittent timescale has also been observed in BH-LMXBs
during their outburst. Worth highlighting is the rich variability of
GRS~1915+105, particularly for its repeatability \citep{Belloni2010}, while less
predictable flaring was seen, e.g., in \vfour \citep[e.g.,][]{alfonsogarzon2018,tetarenko2017}. 
Flaring in the radio band is an established signature of the hard to
soft-state transition of BHB outbursts \citep{fender2004b}. 
On much shorter timescales of $\lesssim\,1\,\mathrm{s}$ flaring and variability
has been attributed to changes in the inner disk region and magnetic reconnection
events above it \citep{lyubarskii1997, uttley2025a}.  
Here we report on a series of exceptionally bright flares
observed with \inte on 2023 July 10. During these flares, which
occurred during a time interval of little more than one hour and
lasted only for $\sim$5--10\,minutes each, the X-ray emission of \cyg
was brighter by a factor of more than 20 compared to the brightest
emission seen in the more than 20\,Ms of \inte observations taken
since its launch in 2002. In Sect.~\ref{sec:flare} we present the
light curves of the flares. We then put the time of the flares in
context of the long-term spectral and state evolution of \cyg in
Sect.~\ref{sec:state} and discuss the detailed behavior of the source
in Sect.~\ref{sec:spec}. We discuss and summarize our results in
Sect.~\ref{sec:dis}.

\begin{figure}
  \resizebox{\hsize}{!}{\includegraphics{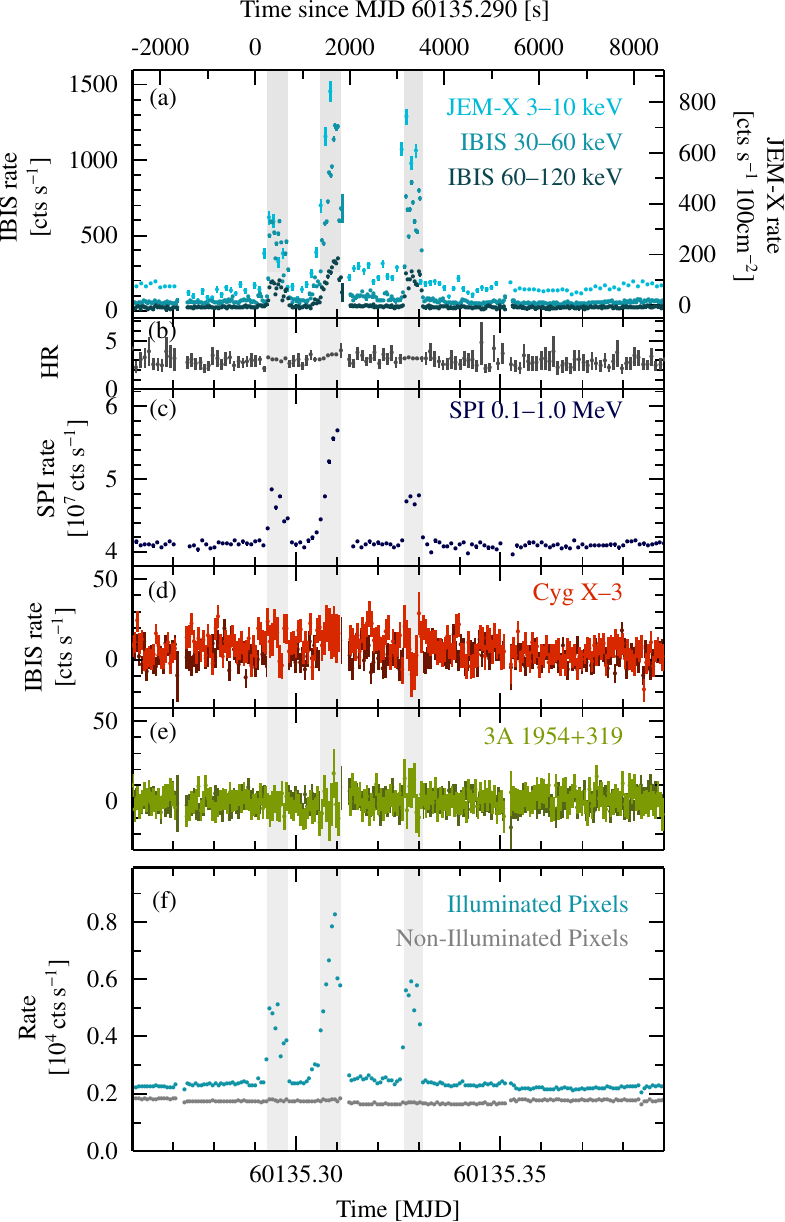}} 
  \caption{Variability of Cyg~X-1 on 2023 July 10. (a) IBIS and
    JEM-X light curves of the flares during rev.~2661. Gray bands
    indicate the time ranges of the flares. (b) Variation of
    the hardness ratio,
    $\mathrm{HR}=(\mathrm{H}-\mathrm{S})/(\mathrm{H}+\mathrm{S})$,
    between the 30--60\,keV and 60--120\,keV \ibis light curves
    (c) Evolution of the total event rate in the SPI detector,
    which is dominated by \cyg. (d) and (e) IBIS light
    curves for the two other sources in the field of view, Cyg~X-3 and
    3A~1954+319, which are unaffected by the flare of \cyg. Lighter
    colors indicate the 30--60\,keV rate, darker ones the 60--120\,keV
    rate. (f) Count rates in those pixels of IBIS illuminated
    and non-illuminated by Cyg X-1. The changes in the base count rate
    around MJD\,60135.27, MJD\,60135.31, and MJD\,60135.35 are due to
    repointing of \inte, resulting in a change in the number of
    (non-)illuminated pixels and of the vignetting of the source.
  } \label{fig:lc}
\end{figure}
 
\begin{figure}
  \resizebox{\hsize}{!}{\includegraphics{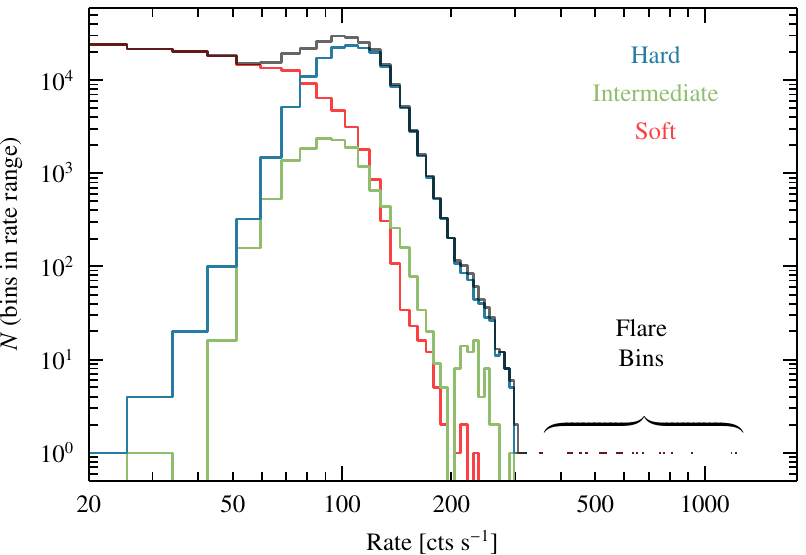}} 
  \caption{Distribution of count rates in 60\,s bins for all \ibis observations
  of Cyg~X-1, separated by state as determined following \citet{Grinberg2013}.
  The bin width of the rate histogram is $80\,\mathrm{cts}\,\mathrm{s}^{-1}$.
  The count rates reached by the flares discussed in this paper are indicated by
  the bracket. At no time before the flares have rates of
  $350\,\mathrm{cts}\,\mathrm{s}^{-1}$ or higher had been reached. The flares
  therefore represent by far the brightest observations ever seen with \ibis for
  \cyg.}\label{fig:histo}
\end{figure}

\section{Strong flaring of Cyg~X-1}\label{sec:flare}

Cyg X-1 has been a regular target of \inte since its launch in 2002
\citep[e.g.,][]{pottschmidt2003b,delsanto2003,CadolleBel2006}. Since 2013 our
team organized regular monitoring observations during the two observing windows
each year, given by visibility constraints, with the aim to further constrain
the hard X-ray polarization found with \inte \citep{Laurent2011, Jourdain2012,
Rodriguez2015}. As part of these regular observations, on 2023 July 10, during
\inte's revolution 2661, during a $<$1\,h long time interval strong flaring was
discovered through the MMODA interface
\citep{neronov2021,ferrigno2022}\footnote{A collection of spectra, images, and
lightcurves of Cyg X-1 obtained with the INTEGRAL telescope is available at
\url{https://www.astro.unige.ch/astroordas/mmoda}.}, with a flux that strongly
exceeded the fluxes detected since the launch of \inte.

For a detailed analysis we extracted spectra and light curves from all the
IBIS/ISGRI \citep{ubertini2003,Lebrun2003} and \jemx \citep{lund2003} data of
\cyg taken during revolution 2661 using Offline Science Analysis (OSA) software
package 11.2. The flares were confirmed in two \inte Science Windows, i.e.,
slightly offset pointings of the \inte satellite\footnote{The science windows
are 266100120010 (MJD 60135.273--60135.311, live time 1611\,s), and 266100130010
(MJD 60135.312--60135.351, live time 1943\,s), with a brief slew between them.
Here and elsewhere in the paper, all MJDs refer to the local \inte satellite
time system and are not barycentered}. Since the two Science Windows have
off-axis angles of $5\fdg6$ and $5\fdg2$, respectively, placing \cyg barely at
the edge of the field of view of \jemx, no spectral analysis with \jemx is
possible. The minute-scale evolution of the flares is shown in
Fig.~\ref{fig:lc}.    The IBIS and JEM-X light curves clearly show the flaring
behavior (Fig.~\ref{fig:lc}a), as does the total event rate measured in the
SPectrometer on \inte \citep[SPI:][see Fig.~\ref{fig:lc}c]{Vedrenne2003}. No
simultaneous information is available in the optical, since \cyg was outside the
${\sim}5^\circ$ field of view of \inte's optical monitor.

In Fig.~\ref{fig:histo} we compare the count rate measured during
60\,s long time bins during the flares to the distribution of the
count rates found in the other ${\sim}330000$ 60\,s-long time bins
covering the ${\sim}23$\,yr of \cyg \inte monitoring (a total exposure
time of ${\sim}20\,$Ms ). The flares clearly represent by far
the brightest events ever seen for \cyg with \inte.

The exceptional brightness of \cyg during the flares makes it important to confirm that
they are not due to some other event in the field of
view, such as a background flare or a gamma-ray burst. First, we check
other sources in the same field for flaring.
The light curves of Cyg~X-3 and 3A~1954+319, the most significantly detected sources in the field of Cyg~X-1, display only a minuscule flux increase during the flaring period (Fig.~\ref{fig:lc}d and e). This slight increase is likely due to effects of the
deconvolution algorithm in the OSA, where a small fraction of the flux
of other sources in the field of view can be misattributed to the
source under study \citep{goldwurm2003}. 

In addition, we generate the light curve of Cyg X-1 without relying on
deconvolution. Since \ibis utilizes a coded mask, for a given pointing direction each source in the field of view only illuminates a
subset of pixels. Figure~\ref{fig:lc}f shows the
count rate measured in those pixels of IBIS that are illuminated by \cyg
(Pixel Illumination Fraction, $\mathrm{PIF}> 0.7$) and compares it with those that are not illuminated.
The only clear signal originates from \cyg itself. No increase is seen in
the pixels that are not illuminated by \cyg. This clearly illustrates that no background activity is the origin of
the flares. Since the likelihood that the flares are from a serendipitous source very close to \cyg is
extremely small, we conclude that the flares must come from \cyg itself.

\begin{figure*}
  \sidecaption
  \includegraphics[width=120mm]{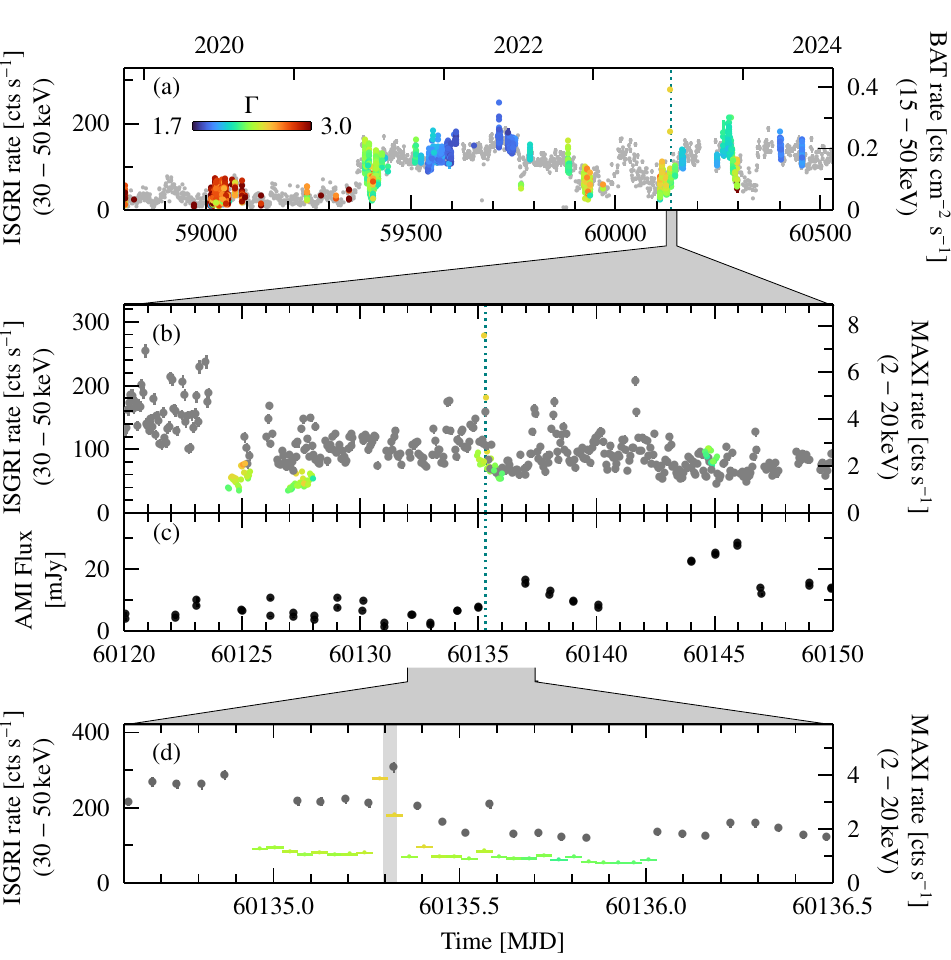}
  \caption{Long-term IBIS and BAT light curve of \cyg. The top panel
    shows the source behavior in the larger context, while the lower
    panels focus on days surrounding the flare. (a) \bat (gray)
    and \ibis count rates, colored by the photon index of the
    corresponding Science Window, determined with applying the
    \texttt{bknpower} model. The canonical state of \cyg derived from
    the \maxi light curves according to \citet{Grinberg2014} in shown
    in the color strip on the top edge of the panel. The time of the
    flaring episode is indicated by the teal, dotted, vertical line.
   (b) IBIS data shown as in (a), but focusing on a shorter time interval and including the \maxi rates, instead of \bat.
    (c) the radio flux density measured with AMI. (d) \maxi
    on-demand data (grey), again together with the IBIS data colored as above). The interval between start of the first and end of the third flare is shaded in light gray and the length of each pointing by the horizontal error bars. 
    A slight increase in the \maxi flux is visible around the
    time of the flare.}\label{fig:cont}
\end{figure*}

Finally, while no pointed observations with instruments on other
spacecraft were performed during the flares, Cyg~X-1 was also
monitored with \maxi, which provides on-demand light curves in user
defined energy bands\footnote{http://maxi.riken.jp/mxondem/index.html}. These light curves are shown together with the more long-term behavior around the flare in Fig.~\ref{fig:cont}. A slight
increase in count rate at the time of the flares is present (Fig.~\ref{fig:cont}d) and might be
attributable to the flare (partially) happening during a \maxi
exposure, which are generally separated by one orbit of about 90\,min.

We therefore conclude that the flares are indeed intrinsic to
\cyg.

\begin{figure}
 
\resizebox{\hsize}{!}{\includegraphics{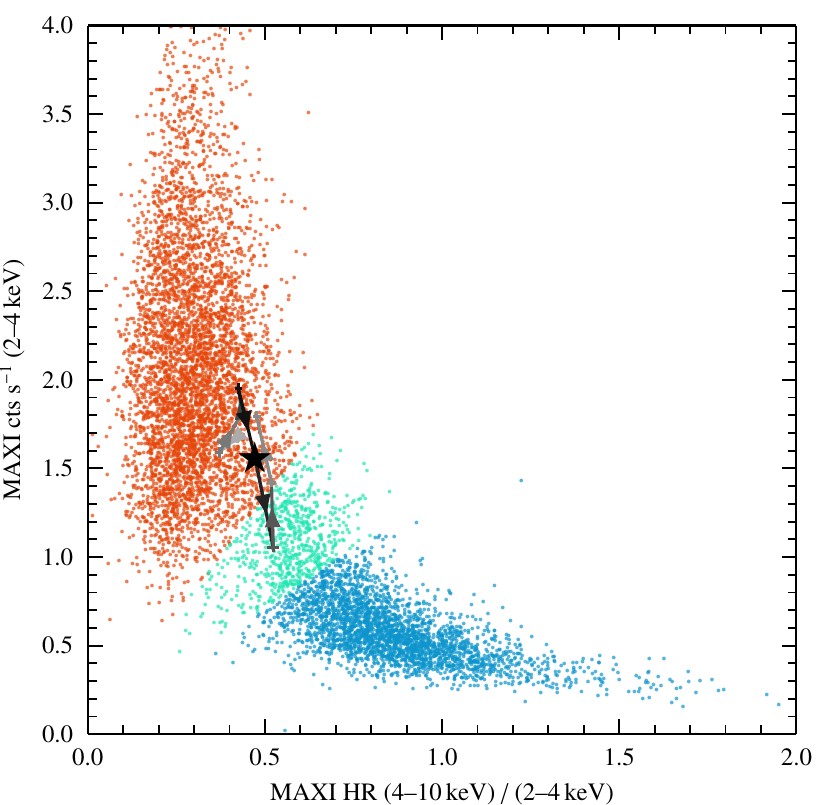}} 
  \caption{Hardness--intensity diagram of \cyg determined from MAXI
    data. The position of \cyg during the flare is indicated by the
    black star. Red, green, and blue dots indicate individual bins in
    the daily light curves of \maxi, classified as soft, intermediate
    or hard state \citep{Grinberg2013}. The source behavior
    four days before and four days after the flaring episode is
    indicated by arrows pointing towards later observations and
    becoming lighter with increased separation from the flare. 
  }\label{fig:state}
\end{figure}

\section{The flaring episode in context} \label{sec:state}

We now turn to the behavior of \cyg in the months surrounding the
flaring episode. As shown in
Fig.~\ref{fig:cont}a, the flares occurred during a phase when the hard
flux of \cyg rose. Such a behavior is commonly seen in this source
during transitions from the soft state to the hard state. Indeed, the
well-established classification by \citet{Grinberg2013}, based on the
MAXI monitor and illustrated in Fig.~\ref{fig:state}, places \cyg
close to the soft-to-intermediate-state transition during the flare.
During the following four daily scans, \cyg briefly entered the
intermediate state.

The general source behavior in the time surrounding the flaring
episode was typical for Cyg~X-1. Reviewing the \inte spectroscopy
during the days surrounding the flare, no strong or sudden changes in
the hard photon index are apparent, implying a stable geometry of the
Comptonizing plasma without any sudden change in flux or hardness
(Fig.~\ref{fig:cont}b).

Some black hole X-ray binaries show increased radio flaring during
such state transitions. To investigate the possibility of a correlated
radio flare, we used data from the Arcminute Microkelvin Imager
\citep[AMI,][]{zwart2008}. The 15\,GHz radio monitoring consists of two
10\,minutes pointings for each day with appropriate observing
conditions, separated by a calibration observation. The radio flux density
per 10\,min pointing and for one linear polarization direction, i.e.,
Stokes I+Q, are shown in Fig.~\ref{fig:cont}c. AMI provided snapshots
shortly before the flaring episode on MJD\,60135.0225 (6.4\,hrs before
the first flare) and then 40.4\,h after the last flare at
MJD\,60137.0145, showing that the flux increased from 7.7\,mJy to
16.5\,mJy, which is within the range of typical variability in the
soft to soft-intermediate state 
 \citep{Rodriguez2015,lubinski_2020}. In the following five days, the radio flux density continuously
decreased. The $\sim$2\,d data gap starting just before the flares
prevents us from making a statement about the presence of a simultaneous
radio flare.

\begin{figure*}
  \resizebox{\hsize}{!}{\includegraphics{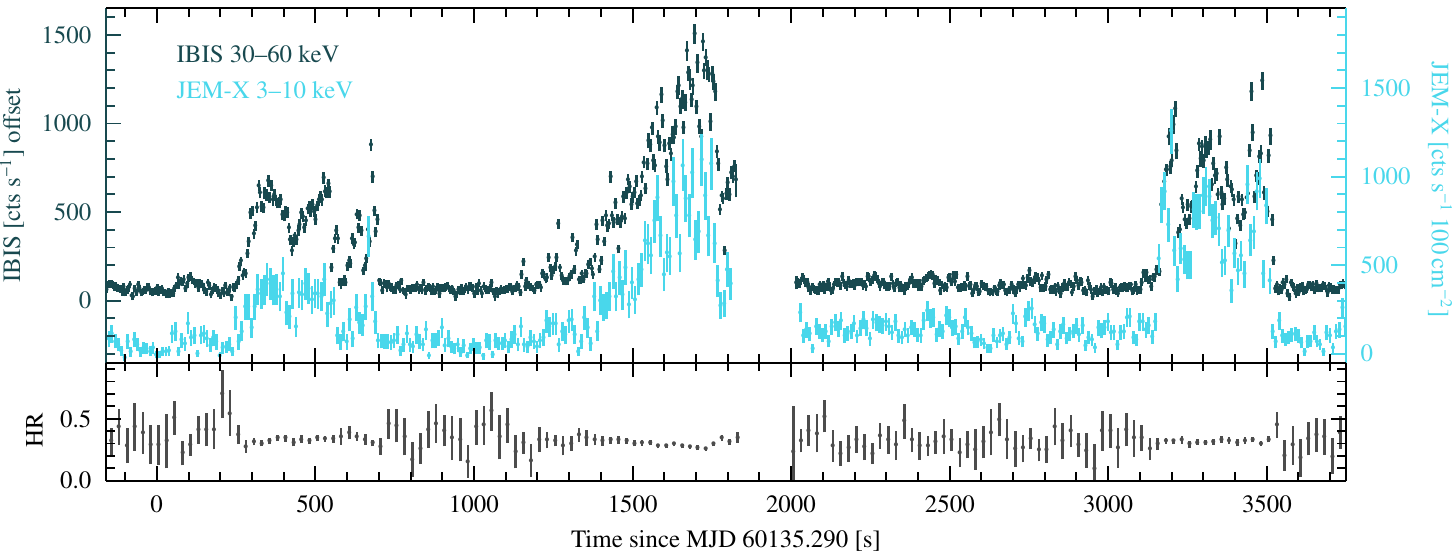}} \caption{Top: JEM-X and
  IBIS light curves of the three flares in the 3--10\,keV and 30--60\,keV band
  with 10\,s and 5\,s time resolution, respectively. The $y$-axis for the IBIS
  data is shifted up for readability. Bottom: Hardness ratio light curve with a
  time resolution of 60\,s. The hardness ratio is defined as
  $\mathrm{HR}=(\mathrm{H}-\mathrm{S})/(\mathrm{H}+\mathrm{S})$ for 30--60\,keV
  and 60--120\,keV. The gap starting around 1900\,s is due to a repointing of
  \inte. }\label{fig:fast}
\end{figure*}

\begin{figure}
  \resizebox{\hsize}{!}{\includegraphics{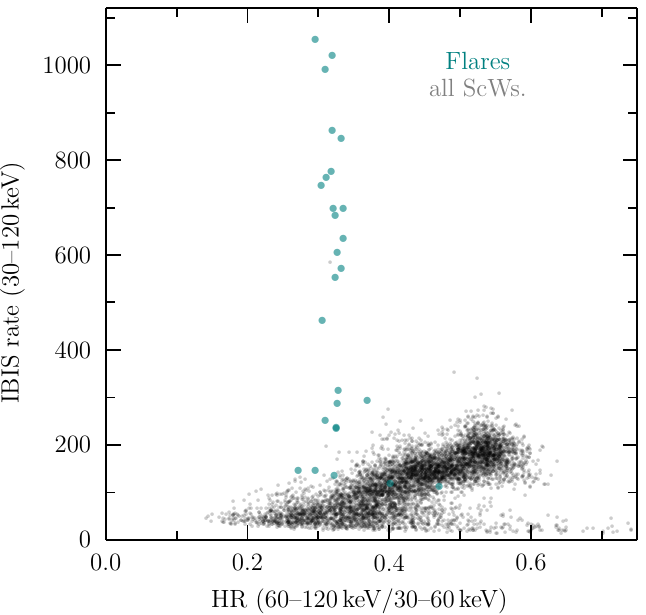}} 
  \caption{Evolution of the hardness ratio and count rate over
    the course of the three outbursts in teal, compared to all other
    \inte observations of \cyg in gray. A bin time of 100\,s
    was used for each data point. No clear trend towards a softening or
    a hardening with increasing flux is visible. }\label{fig:flate}
\end{figure}

\begin{figure*}
  \resizebox{\hsize}{!}{\includegraphics{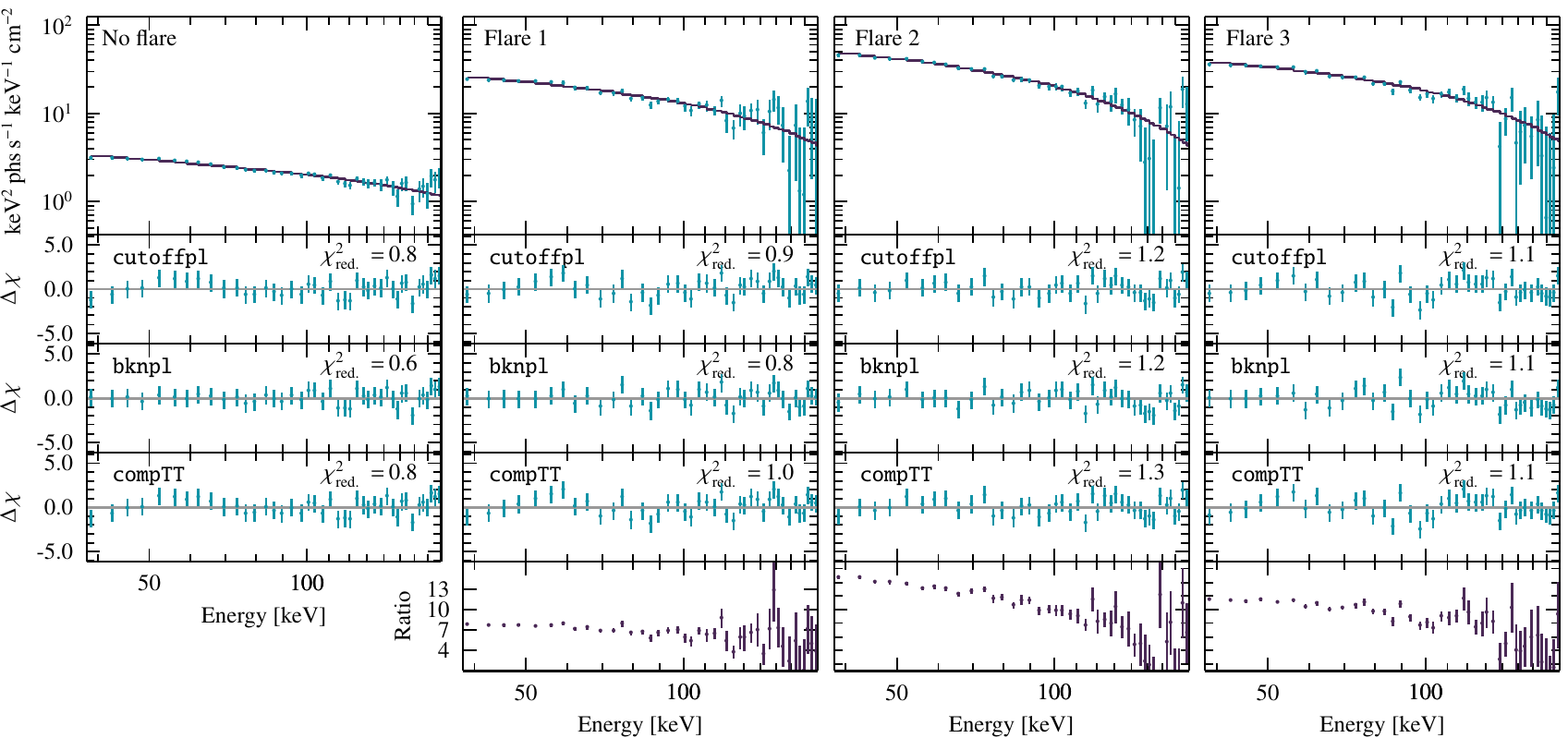}} 
  \caption{IBIS spectra for revolution 2661 with the flare GTIs removed (``no
  flare'') and for the three flares. Top: Unfolded spectra and the best-fit
  \texttt{cutoffpl} continuum. Middle: Residuals between the data and the
  best-fit models of Table~\ref{tab:fits}. Bottom: Ratio between each flare
  spectrum and the no flare spectrum.}\label{fig:spec}
\end{figure*}

\section{Source behavior during the flares} \label{sec:spec}

Figure~\ref{fig:fast} shows a zoom-in onto the IBIS light curves at
5\,s resolution. The first and third flare are characterized by very
fast rises of 30--50\,s in duration, with very fast doubling
timescales of 15\,s, followed by a ``flat top'' with an average IBIS
count rate of ${\sim}500\,\mathrm{cts}\,\mathrm{s}^{-1}$, but with
very strong normalized rms variability on timescales of tens of
seconds, 51\% and 36\% for flare 1 and 3 at a time resolution of
$\Delta t=10$\,s, and slightly less at a time resolution of 20\,s. The
variability is significantly increased with respect to the 24\% at
$\Delta t=10$\,s we measure outside the flares\footnote{The
normalized rms-variability is only an approximate estimator for the
variability in the case of red noise light curves \citep{vaughan2003},
but unfortunately the shortness of the flares precludes a more
detailed characterization of the variability properties, e.g., through
power spectra.}. At the end of the flares \cyg quickly returns
to its previous count rate, and the variability goes back to its
pre-flare value.

In contrast, the second flare is characterized by a sudden increase in
normalized rms variability (39\% at $\Delta t=10$\,s) and a much
slower rise (doubling time scale: ${\sim}150\,\mathrm{s}$) to a peak count rate of %
about $1400\,\mathrm{cts}\,\mathrm{s}^{-1}$. Unfortunately, the
decay back to the pre-flare luminosity is only partly covered due to a
spacecraft slew, but overall the shape of the second flare appears to
be almost triangular or Gaussian in shape. Again, with 36\% the rms
variability during this episode is much stronger than outside of the
event.

As discussed in Sect.~\ref{sec:state}, at timescales of hours to years,
flux changes in Cyg~X-1 are closely connected with spectral changes,
as the source moves from the hard to the soft state and back. We
illustrate this in the hardness-intensity-diagram of Cyg~X-1 shown in
Fig.~\ref{fig:flate}, which is based on 100\,s resolution data from
all 22\,years of \inte-monitoring of the source. In contrast,
despite the large flux amplitude during the flares, the spectral
hardness remains almost constant during these episodes
(Fig.~\ref{fig:flate}, teal data points and see also Fig.~\ref{fig:lc}
for a light curve of the hardness ratio).

To further quantify possible changes of the spectral shape, we
extracted IBIS spectra from each of the flares as well as a spectrum
for the time outside the flares (defined as the time interval
outside the shaded regions in Fig.~\ref{fig:lc}). Due to the
off-axis angle of $> 5^\circ$ during the flare, we did not use the
\jemx data for spectral analysis. To take into account calibration
uncertainties, a systematic uncertainty of 3\% of the count rate was
added in quadrature to the statistical uncertainty of each spectral
bin.

\begin{table} \centering
  \caption{Best-fit parameters for the IBIS spectra of the time outside of the flares and the three flares.}\label{tab:fits}
  {\small
\renewcommand{\arraystretch}{1.2}
  \begin{tabular}{lrrrr}
  \hline\hline
  & Non-Flare&Flare 1&Flare 2&Flare 3 \\
  \hline
  $T_\mathrm{exp}$  & 57488 & 302 & 260 & 242 \\\hline
    \multicolumn{5}{l}{\texttt{cutoffpl}}     \\
    $\Gamma$ & $2.17^{+0.19}_{-0.20}$ & $1.7\pm0.4$ & $1.5\pm0.4$ & $1.6\pm0.4$ \\
    $E_\mathrm{fold}$ & $190^{+170}_{-70}$ & $65^{+31}_{-17}$ & $47^{+13}_{-9}$ & $53^{+18}_{-12}$ \\
    $F_{40-80}$ & $1.88\pm0.04$ & $14.2\pm0.4$ & $25.5\pm0.6$ & $33.3\pm0.8$  \\
 $\chi^2/\mathrm{dof}$  & 46.2/59 &  55.7/59 & 71.4/59 &  63.4/59 \\ \hline
     \multicolumn{5}{l}{\texttt{brokenpl}}     \\
     $\Gamma_1$ & $2.17^{+0.17}_{-0.43}$ & $2.13^{+0.23}_{-0.31}$ & $2.44^{+0.17}_{-0.23}$ & $2.21^{+0.25}_{-0.33}$ \\
     $E_\mathrm{break}$ & $55\pm9$ & $55^{+7}_{-6}$ & $64\pm8$ & $57^{+11}_{-7}$ \\
     $\Gamma_2$  & $2.70\pm0.08$ & $3.06^{+0.15}_{-0.13}$ & $3.49^{+0.23}_{-0.17}$ & $3.21^{+0.20}_{-0.13}$ \\
     $F_{40-80}$ & $1.92\pm0.05$ & $14.5\pm0.4$ & $25.7\pm0.6$ & $33.7\pm0.9$ \\
 $\chi^2/\mathrm{dof}$ &36.2/58 &  49.2/58 & 69.1/58 &  62.7/58  \\ \hline
    \multicolumn{5}{l}{\texttt{compTT}}     \\
    $kT_\mathrm{e}$  & $ 130^{+170}_{-70}$ & $33^{+22}_{-8}$ & $25^{+6}_{-4}$ & $28^{+8}_{-5}$ \\
    $\tau$  & $0.10^{+0.28}_{-0.10}$ & $0.9^{+0.4}_{-0.5}$ & $1.1\pm0.4$ & $1.1\pm0.4$ \\
    $F_{40-80}$ & $1.88\pm0.04$ & $14.1\pm0.4$ & $25.4\pm0.6$ & $33.1\pm0.8$ \\
 $\chi^2/\mathrm{dof}$ & 46.2/59 &  57.5/59 & 74.8/59 &  65.9/59 \\
\hline
  \end{tabular}
  \tablefoot{ Uncertainties are at the 90\% level for one parameter of
    interest. Exposure time , $T_\mathrm{exp}$ is given in s, fluxes, $F_{40-80}$, are in units of
    $10^{-9}\,\mathrm{erg}\,\mathrm{s}^{-1}\,\mathrm{cm}^{-2}$ for the
    40--80\,keV band, the break and folding energies,
    $E_\mathrm{break}$ and $E_\mathrm{fold}$, as well as the
    temperature of the Comptonizing plasma, $kT_\mathrm{e}$, are given in 
    keV. The seed photon temperature for the \texttt{comptt}-model,
    where a disk geometry is assumed, was fixed at 200\,eV.} }
\end{table}

We modeled the four spectra with typical models for X-ray spectra of
black hole X-ray binaries in the hard X-rays: a simple powerlaw, an exponential cutoff power
law (\texttt{cutoffpl}), a broken power law (\texttt{bknpow}), and
thermal Comptonization \citep[\texttt{comptt};][]{Titarchuk_1994a}. The latter
three are all able to describe a break or cutoff in the powerlaw
spectrum.  The best fitting parameters are listed in Table~\ref{tab:fits}. All
uncertainties are
given at the 90\% level for one parameter of interest, and we employed
ISIS\footnote{https://space.mit.edu/cxc/isis/} for spectral fitting. The simple
power law is unable to provide a satisfactory fit, leading to $\chi^2/\mathrm{dof}$ of $88.0/60$, $145.7/60$, and $120.3/60$  for the three flares, respectively. 
The three models with an intrinsic turnover or break, those shown in Fig.~\ref{fig:spec}, all
describe the spectra well. 

This is in agreement with previous work,
which has shown the break in the \texttt{bknpow} model to describe the spectra of BHBs
similarly well to the cutoff powerlaw and even more complex jet models
\citep{markoff2003,Nowak_2011a}.  

A direct comparison between the
non-flare and flare spectra, determined by dividing them
by each other, shows that the flare spectra are slightly softer
(Fig.~\ref{fig:spec}, bottom panel). In the \texttt{cutoffpl}-fits
this softening is reflected by a decrease of the folding energy, similarly, while
the \texttt{comptt} models describe the softening with a decrease in the temperature of the
Comptonizing plasma, $kT_\mathrm{e}$, together with a strong increase
in the optical depth (see Table~\ref{tab:fits}). We caution, however, that the data are not good
enough to distinguish between these different models, and that the
energy coverage of IBIS alone is not sufficient to separate the
spectral continuum from the relativistic reflection hump that
contributes significantly in the IBIS-band. A direct physical
interpretation of the spectral parameters is therefore unadvisable.

Using the \texttt{comptt} model,
 the peak 1--100\,keV luminosities are
$1.1\times 10^{38}\,\mathrm{erg}\,\mathrm{s}^{-1}$,
$2.6\times 10^{38}\,\mathrm{erg}\,\mathrm{s}^{-1}$,
$2.2\times 10^{38}\,\mathrm{erg}\,\mathrm{s}^{-1}$, for flare 1 through 3, while the
total 1--100\,keV fluence contained in the three flares
is $3.01\,\times 10^{40}\,\mathrm{erg}$, $4.74\,\times
10^{40}\,\mathrm{erg}$, and $3.84\,\times 10^{40}\,\mathrm{erg}$.
 As these luminosities rely on the extrapolation of the spectral fits towards lower energies, they are only a rough lower limit. Even though we do not have direct spectral data, however, we can infer from the \jemx that  the soft emission likely shows an increase similar to the increase of the 
 luminosities in the 30--100\,keV range, which are
$2.3\times 10^{37}\,\mathrm{erg}\,\mathrm{s}^{-1}$,
$4.3\times 10^{37}\,\mathrm{erg}\,\mathrm{s}^{-1}$,
$3.5\times 10^{37}\,\mathrm{erg}\,\mathrm{s}^{-1}$, for the three flares, respectively.

The hard photon index $\Gamma_1 \sim 2.1 \textrm{--} 2.4$, inferred from the \texttt{brokenpl} fits, during the
three flares and the surrounding observation matches what has been seen in
\cyg during the soft to intermediate state
\citep[e.g.,][]{Nowak_2011a,lubinski_2020}. It is remarkable that the
spectra during the flares, especially flare~1 and flare~3, show very
little change in spectral slope. Only the spectra of flare~2 display mild
softening. This becomes visible
through the ratio with the non-flare spectrum, which decrease towards higher
energies, as seen in Fig.~\ref{fig:spec}. This contrasts with the rather
vertical evolution in the hardness-intensity diagram of Fig.~\ref{fig:flate},
but might result from the chosen energy bands and time binning. 

\section{Discussion and Conclusions} \label{sec:dis}

Before we discuss the possible physical origin of the flaring, we
briefly summarize their main properties:
\begin{itemize}
\item  the three flares represent extreme source behavior that has not
  been seen previously in the 21\,years of monitoring of \cyg with
  \inte, nor in the earlier \rxte monitoring between 1997 and 2012,
\item the flares occurred during the soft-intermediate state, when
  Cyg~X-1 was moving towards the hard state,
\item the flares occurred at orbital phase $\phi_\mathrm{orb}=0.01$
  based on the ephemeris of \citet{brocksopp1999}, i.e., close to upper conjunction of the black hole,
\item the flares have peak luminosities of 1--100\,keV luminosity of $1.1\textrm{--} 2.6\times 10^{38}\,\mathrm{erg}\,\mathrm{s}^{-1}$ ($4.1\%\textrm{--} 9.7\%\,L_\mathrm{Edd}$), a
     dynamic flux range of ${\sim}15$, and a duration of about 400\,s
  each, with fluences of $3\textrm{--} 5\times 10^{40}$\,erg each,
\item the intensity profiles are complex with fast rise and decay times
  ${\sim}10\,\mathrm{s}$ for the first and third flare, and a slow rise and fast decay for the second flare,
\item during all three flares the normalized rms variability is significantly increased,
\item there is little spectral change in the hard X-rays, with only a
  slight softening $>30\,\mathrm{keV}$ (see ratio panel in Fig.~\ref{fig:spec}). 
\end{itemize} 
Besides their timing properties mentioned above, it is the extreme peak luminosities that make
the flares stand out above the typical variability seen in \cyg. To our
knowledge, the only comparable event in Cyg~X-1 was detected by \rxte in 2005
April \citep{wilms2007a}. This event had a similar duration of ${\sim}
10\,\mathrm{min}$ and coincided with a radio flare that was delayed by about
400\,s with respect to the X-ray flare.
Similarly to the flares observed by \inte, \cyg was in the intermediate
state, transitioning towards the hard state. A notable difference is
the lower dynamic range: while we observe a ${\sim}15\times$ increase
in hard X-ray flux, \citet{wilms2007a} saw the radio and X-ray flux increase by only about
a factor of three\footnote{Since the PCA and IBIS energy ranges differ
significantly, a more precise comparison of the fluxes is difficult,
in addition, the peak X-ray flux of the \rxte flare could be
underestimated as \rxte only caught its decay.}. Radio flares with
behavior similar to that studied by \citet{wilms2007a} were also
discussed by \citet{fender2006}, including a strong 140\,mJy radio
flare with a duration of about one hour during the intermediate state
of \cyg, but unfortunately these radio flares lacked simultaneous
X-ray data.
The orbital phase aligns closely with phase zero. However, we would primarily
expect the absorption to vary as a line-of-sight effect \citep{grinberg2015,
szostek2007} and not the phenomena related to the accretion flow itself. For
comparison, the flare reported in \citet{wilms2007a} occurred around orbital
phase 0.82. Concerning the spectral shape, the necessity of a cutoff does point
more towards thermal origin of the X-ray emission, rather
than synchrotron emission. Jet models such as those discussed by
\citet{Markoff2005, Nowak_2011a, maitra2017a, kantzas2021} can, however, produce
spectral shapes more complex than a simple powerlaw. A jet origin for the
observed flares can therefore not be excluded based on spectral shape. 

Before investigating the possible origin of the flares, we emphasize that the
flares discussed here are different from shorter-term variability in accreting
black holes, which is sometimes explained in terms of shot-noise
\citep[e.g.,][]{bhargava2022a,gierlinski2003}. The latter is thought to be
connected to magnetic reconnection and the evolution of plasmoids in a current
sheet above the disk \citep[e.g.,][]{ripperda2020,elmellah2022,merloni2001} and occurs on
timescales of $10\textrm{--}100\,r_\mathrm{g}/c$. The variability of $\sim$10\,min
seen here, however, corresponds to $10^6\,r_\mathrm{g}/c$, or approximately the light
crossing time between the donor and the black hole\footnote{We caution that the
term ``flare'' is not well-defined in astronomy, and may denote events on very
different timescales. For black hole X-ray binaries alone, ``flare'' has been
used for large amplitude flux changes on time scales of seconds, minutes, days,
and months, which are likely to be due to very different physics. In the
following, we use the term solely for flux changes on timescales less than a few
100\,minutes.}. 
Conversely, following a simple scaling by mass as appropriate for black holes, the minute-scale variability as seen in AGNs corresponds to millisecond variability in BHBs. 

Such short-term variability has previously also been attributed to the innermost
region of the disk. Variable seed photons from the disk can then be upscattered in the
corona to induce variability in harder energy-bands \citep[][and references therein]{uttley2025a}.
In such a scenario, the dynamic timescale is, however, of the order of $10^{-3}$\,s, comparable to
the Keplerian timescale, and therefore much
faster than the observed flaring episodes \citep{lyubarskii1997, done2007}. An overview of
the different timescales in accreting black holes is given by \citet{kara2025}. 
If the variability originates from the disk, it must therefore have its origin in the outer regions, where the Keplerian and viscous timescales are much longer. 
In such a case, the model of propagating fluctuations could be considered a possible explanation for the flares.
Under this assumption, aperiodic fluctuations in the accretion are introduced, classically by turbulent changes
in the disk viscosity at all radii of the disk. These fluctuations propagate inwards towards the
hotter region of the accretion disk, reproducing the observed time lags and power spectral density  \citep[see, e.g.,][]{lyubarskii1997, uttley2014,ingram2011}.
A sudden increase in the mass accretion rate originating at a radius of ${\sim}10^2\,r_\mathrm{g}$ could
lead to variability on the order of 100\,s, corresponding to the viscous time scale. The inwards propagation is, however, a diffusion process
and smoothens out the induced variability.  The rapid rise and decay time of the flares are therefore
atypical for propagating fluctuations. The model is further usually only applicable to
stochastic variability, not individual flares. Due to a lack of soft spectra and light curves, however, such a scenario cannot be fully ruled
out.

Similarly, an explanation though the dissipation time $t_\mathrm{D}$ after
magnetic heating, as used by \citet{merloni2001} following earlier work by \citet{haardt1994} requires a region size of
\begin{equation}
 \dfrac{r}{r_\mathrm{g}}  \simeq 1.6 \times 10^4  \left(\dfrac{ b}{30}\right)^{-1} \left(\dfrac{ M_\mathrm{BH}}{21 M_\odot}\right)^{-1}  \dfrac{t_\mathrm{D}}{100\,\mathrm{s}}  ,
\end{equation}
with the unitless dissipation velocity $b=c/v_\mathrm{diss}$. This is much larger than the inner
regions of the accretion disk. Accretion fluctuations alone are therefore
disfavored as explanations, but could still serve as triggers for downstream
changes in the jet or corona.

For examples of flaring on times scales of hours we can expand our view
to other BHB systems beside \cyg. Flaring has been observed in the radio band
for several transient black hole X-ray binaries, where they occur around state
transitions and may be correlated with changes in the X-ray timing behavior
\citep[e.g.,][]{fender2009,millerjones2012}. This has relevance to our
discussion as these have also been attributed to ejection events through the jet.
As these state transitions, however, occur over days, they are therefore phenomena on very
different timescales and the physical mechanism behind these flares could be
very different. Direct observations of such hour-long radio flares are rare.
\citet{homan2020a} describe a strong radio flare in MAXI~J1820+070, lasting for
$\sim$2.5\,h and reaching $\sim$50\,mJy, which was accompanied by a
quasi-simultaneous weak flare in the 7--12\,keV \nicer band. In this case,
however, the X-ray flux only increased by a few percent. Flaring on a timescale
of minutes in the optical and X-rays has also been studied for the 2015 outburst
of \vfour \citep{Rodriguez2015,tetarenko2017,alfonsogarzon2018}. Here, these
flares display a similar complexity in shape to those seen in Cyg~X-1, albeit at
a much lower dynamic range. 
Explanations for the type of minute-long flares discussed here build on the observation
that, for those flares where simultaneous broad-band data exist, flaring is not
limited to the X-rays alone, but seen across the electromagnetic spectrum. Time
delays are common, where longer wavelengths are delayed with respect to shorter
ones \citep[e.g.,][and references therein]{fender2023}. As suggested by
\citet{fender2004b} and \citet{wilms2007a}, a possible model explaining these
events is some kind of ejection event, similar to the flaring seen in blazars
first discussed by \citet{vanderlaan:1966}.
In such models, a bubble of relativistic electrons is ejected from the accretion
flow and subsequently cools down, moving the peak of the emission to longer
wavelengths. The slight spectral changes seen in the hard X-rays are interpreted
as a change of the Compton-$y$-parameter due to the expansion and cooling of the
ejected material. See \citet{younsi2015} for detailed computations of plasmoid
ejections that take general relativistic effects into account and suggest that
strong variability and multiple reflares on timescales of tens of
$r_\mathrm{g}/c$ are in principle possible.
Such ejections have been invoked for the explanation of flares in the lightcurves of \vfour with timescales  of
${\sim}200\,\mathrm{s}$ by \citet{maitra2017a} on the basis of multiwavelength data. In this case, a
predominant disk origin was explicitly ruled out on the basis of correlated optical variability.
The very characteristic heart-beat variability and particularly large orbital period, and therefore
disc make \vfour a unique system \citep{fender2004b, steeghs2013}. This might limit to which we can
transfer explanation for flaring in \vfour to \cyg, as the disc might behave very differently. Yet
we hold that close to the black-hole the physics around the black hole should still be similar and
\vfour serves as a valuable comparison.  

As suggested by \citet{rodrigue15_V404} for \vfour, the amplitude of the
flaring could be increased due to directly boosted radiation when the jet axis
is pointed directly towards the observer. Such a boosting would imply a fairly
large deviation of the jet direction during the flares with respect to its
default orientation. The nominal angle between the jet axis (and the orbital
angular momentum) and our line of sight at upper conjunction
($\phi_\mathrm{orb}=0$) is ${\sim}60^\circ$
\citep{Krawczynski2022,MillerJones2021,Orosz2011}. Such a large deviation is
difficult to achieve. Models for the warping of the inner accretion disk in
Cyg~X-1 imply a warp of only ${\sim}30^\circ$ \citep{ibragimov2007} in order to
explain the $\sim$150\,d or $\sim$294\,d superorbital variability in the system
\citep{zdz2011,brocksopp1999,priedhorsky1983,kemp1983} and the optical
polarization variability \citep[][and references therein]{kravtsov2023}. It is
therefore unlikely that regular disk warping and variable boosting is the sole
source of variability.  
An alternative explanation for the flares is the interaction of structures in the stellar wind of the high-mass donor star and the jet. Based on numerical modeling of clump-jet interaction, \citet{perucho2012} show that for winds with a steep clump size distribution, it is possible for large clumps to enter the jet and be completely disrupted, potentially even choking off the jet. The simulations by \citeauthor{perucho2012} and \citet{araudo2009} predict rare, but luminous flares with durations of $1000\,\mathrm{s}\, (R_\mathrm{clump}/10^{11}\,\mathrm{cm}) (c_\mathrm{c}/10^8\,\mathrm{cm})$, where $R_\mathrm{clump}$ is the clump radius and $c_\mathrm{clump}= (2 L_\mathrm{jet}/\pi v_\mathrm{jet} \rho_\mathrm{clump} R_\mathrm{jet}^2)^{1/2}$ is the speed of sound in the clump. Here, $v_\mathrm{jet}$ is the speed of the jet, $L_\mathrm{jet}$ its luminosity, and $R_\mathrm{jet}$ its radius at the point of interaction, while $\rho_\mathrm{clump}$ is the density of the clump. Observations of X-ray dips in Cyg~X-1 show that the wind of its donor, HDE226868, is clumpy \citep[][and references therein]{hirsch2019,grinberg2015}. Line-of-sight variations of the absorbing column are stronger close to upper conjunction of the black hole \citep{grinberg2015}, indicating that more clumps pass through our line of sight, probably close to the black hole. This is exactly the time interval when the flare was observed, although this is probably a coincidence. We note, however, that if the flare is due to a clump destruction event it is puzzling why its spectral shape would be similar to that seen outside the flare, and why we see three flares and not just one. In addition, while clumps are normal in high-mass stellar winds \citep[][and references therein]{owocki1988}, with a filling factor of $\sim$11\% for Cyg~X-1 \citep{rahoui2011}, they are not expected to exist in low-mass X-ray binaries, where flaring is also observed. 
To conclude, we serendipitously observed an unprecedented set of
${\sim}$10\,minute long, X-ray flares with fluences in excess of
$10^{40}$\,erg, with hard X-ray luminosities that are at least three
times higher than anything seen before in the $\ge 22\,$years of \inte
monitoring of \cyg. The flares occurred in a state where the accretion
flow of black hole binaries is hypothesized to be unstable. The strong flaring might be related to (i) some kind of ejection event, (ii) a restructuring of the outflow (``jet'') in the system, or (iii) the interaction of a clump in the stellar wind with the jet. The observations presented here illustrate the need for continued monitoring even of supposedly ``well-known'' sources, since it allows us to catch dramatic and very rare events in such systems.
\begin{acknowledgements}
We especially acknowledge the crucial contribution of Katja Pottschmidt -- not only to this
paper but the field of Black-hole timing in general. Without her support, mentorship, and scientific insight
this work would not have been possible. Her untimely passing is felt sorely.
This work has been partially funded by the Bundesministerium f\"ur
Wirtschaft und Klimaschutz under Deutsches Zentrum f\"ur Luft- und
Raumfahrt grant 50\,OR\,1909. 
This research is supported by the DFG research unit FOR 5195 `Relativistic Jets in Active Galaxies' (project number 
443220636, grant number WI 1860/20-1).
TB \& JR acknowledge partial funding
from the French Space Agency (CNES). The material is based upon work
supported by NASA under award number 80GSFC24M0006. MP acknowledges support by the Spanish Ministry of Science trough Grant \texttt{PID2022-136828NB-C43}, and by the Generalitat Valenciana through grant \texttt{CIPROM/2022/49}. The research is
based on observations with \inte, an ESA project with instruments and
science data center funded by ESA member states (especially the PI
countries: Denmark, France, Germany, Italy, Switzerland, Spain) and
with the participation of Russia and the USA.
This research has made use ISIS 1.6.2-51 \citep{Houck2000} and of ISIS functions (ISISscripts) provided by ECAP/Remeis observatory and MIT (\url{https://www.sternwarte.uni-erlangen.de/isis/}).
\end{acknowledgements}
\bibliographystyle{aa} 
\bibliography{references}
\end{document}